\documentclass[a4paper]{revtex4}
\usepackage{textcomp}
\usepackage[english]{babel}
\usepackage{graphicx}
\usepackage{rotating}
\usepackage{amsmath}
\usepackage{makeidx}
\usepackage{amsfonts}
\usepackage[ansinew]{inputenc}
\usepackage[usenames,dvipsnames]{pstricks}
\usepackage{epsfig}
\usepackage{pst-grad} 
\usepackage{pst-plot} 
\usepackage[colorlinks,hyperindex]{hyperref}
\usepackage[font=footnotesize,caption=false]{subfig}

\newcommand{\beq}{\begin{equation}}
\newcommand{\eeq}{\end{equation}}

\newcommand{\bea}{\begin{eqnarray}}
\newcommand{\eea}{\end{eqnarray}}

\makeindex

\begin{document}

\title{Unraveling the ``Green Gap'' problem: The role of random alloy fluctuations in InGaN/GaN light emitting diodes}
 
\author{Matthias Auf der Maur$^*$}
 \affiliation{Department of Electronic Engineering, University of Tor Vergata,
 Via del Politecnico 1, 00133 Rome, Italy}
\email[]{auf.der.maur@ing.uniroma2.it}
 \author{Alessandro Pecchia}
\affiliation{CNR-ISMN, via Salaria Km. 29.300, 00017 Monterotondo, Rome, Italy}
\author{Gabriele Penazzi}
\affiliation{Bremen Center for Computational Materials Science, University of Bremen,
28359 Bremen, Germany}
\author{Walter Rodrigues}
\author{Aldo Di Carlo}
 \affiliation{Department of Electronic Engineering, University of Tor Vergata,
 Via del Politecnico 1, 00133 Rome, Italy}



\begin{abstract}
White light emitting diodes based on III-nitride InGaN/GaN quantum wells currently offer the highest overall efficiency for solid state lighting applications.
Although current phosphor-converted white LEDs have high electricity-to-light conversion efficiencies, it has been recently pointed out that the full potential of solid state lighting could be exploited only by color mixing approaches without employing phosphor-based wavelength conversion.
Such an approach requires direct emitting LEDs of different colors, in particular in the green/yellow range ov the visible spectrum.
This range, however, suffers from a systematic drop in efficiency, known as the ``green gap'', whose physical origin has not been understood completely so far.
In this work we show by atomistic simulations that a consistent part of the ``green gap'' in c-plane InGaN/GaN based light emitting diodes may be attributed to a decrease in the radiative recombination coefficient with increasing Indium content due to random fluctuations of the Indium concentration naturally present in any InGaN alloy.


\end{abstract}

\maketitle



III-nitride based light emitting diodes (LEDs) have become since their breakthrough in 1993~\cite{Nakamura1993} the most promising candidates for ultra-high efficiency solid state lighting (SSL), earning S. Nakamura, H. Amano and I. Akasaki last year's Noble prize in physics.
Currently, the power conversion efficiency of blue InGaN/GaN LEDs is exceeding 80\%~\cite{Narukawa2010}, and wide adoption of ultra-efficient SSL solutions would allow for substantial increase in energetic efficiency of generic lighting and to potential energy saving~\cite{AdOM_Tsao_2014}.
Usually, white light emission is obtained by partially converting the emission of a blue LED to the yellow/green spectral range by means of a phosphor.
This conversion is associated with an energy loss known as Stokes' loss, which is in the order of 25\% and therefore limits the highest attainable white phosphor-converted LED (PC-LED) efficiency to well below 100\%~\cite{AdOM_Tsao_2014}.
This loss mechanism can be eluded by eliminating the phosphor-based down-conversion and using direct color mixing instead, combining the light of several LEDs emitting at different wavelengths, usually blue, green, red, and possibly yellow.
In fact it has been pointed out recently, that in order to exploit the full potential of SSL it will be necessary to eventually eliminate the phosphor-based down-conversion by moving to color mixing approaches based on semiconductor-only multi-color electroluminescence~\cite{AdOM_Tsao_2014,SREP_Jeong_2015,ODonnell}.
This will allow for the highest possible efficiency in light generation, described by the internal quantum efficiency (IQE), and for smart-lighting applications requiring particular emission spectra or detailed control on color mixing~\cite{AdOM_Tsao_2014} and for increased lifetime.

\begin{figure}[hbt]
\centering
\includegraphics[width=10cm]{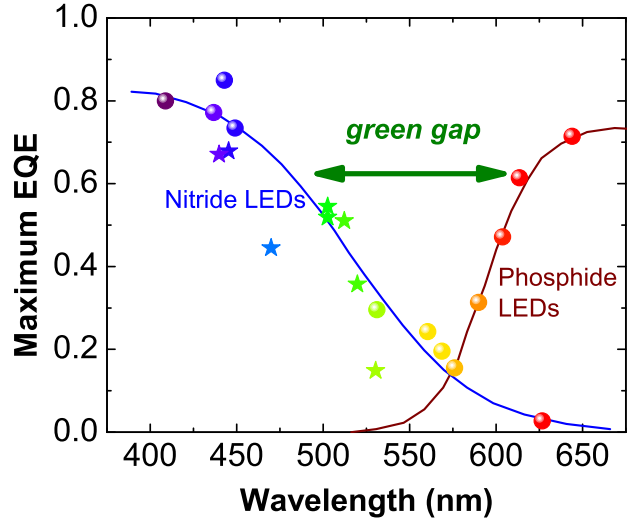}
\caption{{\bf The green gap.} Maximum external quantum efficiency (EQE) of different commercial nitride and phosphide LEDs (spheres), illustrating the ``green gap'' problem. Data points have been taken from~\cite{Karpov_compsem_2015}. The lines are guides to the eye. The stars give the EQE of the nitride single quantum well LEDs from~\cite{Schiavon} we have used for comparison with simulations.}
\label{greengap}
\end{figure}

The most important issue hampering the transition to phosphor-free solutions is the ``green gap''.
The ``green gap'' indicates a severe drop in efficiency of green/yellow emitters compared to blue and red ones~\cite{Taylor2012}, for III-nitride and III-phosphide technology, respectively, as shown in Fig.~\ref{greengap}.
The low efficiency of green LEDs is particularly critical, since phosphor-free white LEDs based on color mixing require at least a green emitter with wavelength around 530 nm~\cite{Karpov_compsem_2015}, lying nearly at the center of the ``green gap''.
Therefore, the efficiency of an all-semiconductor white light source is limited by the efficiency of the green emitter, and in fact today's white PC-LEDs have higher efficiency than green LEDs~\cite{AdOM_Tsao_2014}.
As a consequence, to reach ultra-efficient white LEDs it is of paramount importance to understand the origins of the ``green gap''.

The most advanced technology for green LEDs is based on c-plane InGaN/GaN multi-quantum-well (MQW) LEDs, as used for blue LEDs but with higher Indium content in the InGaN quantum wells (QWs).
To understand the ``green gap'', it is therefore necessary to study the changes in device performance with increasing Indium content.

It is known that both decreasing material quality and increasing quantum confined Stark effect (QCSE) due to the high polarization fields in c-plane InGaN/GaN QWs reduce LED efficiency at increasing Indium content~\cite{Zhao_2013,Schiavon}.
The former leads to increased non-radiative recombination, whereas the latter reduces the overlap between electron and hole wavefunctions and thus the associated momentum matrix elements.
Recent experimental results suggest however, that the decrease of the radiative recombination coefficient with increasing Indium content is stronger than expected from QCSE alone~\cite{Schiavon}.
A possible explanation for the additional decrease in radiative recombination rate could be localization of electrons and holes in the QW plane due to statistical fluctuations in the InGaN alloy~\cite{AufderMaur2014_rnd,Schulz2014}.
In fact it is known that even in a uniform alloy random fluctuations of the local Indium concentration lead to statistical spread of the electronic states's energies and partial localization of the wavefunctions, when calculated with atomistic models like tight-binding~\cite{Lopez_2014,Caro2013,Schulz2014}.

\begin{figure}[hbt]
\centering
\includegraphics[width=\columnwidth]{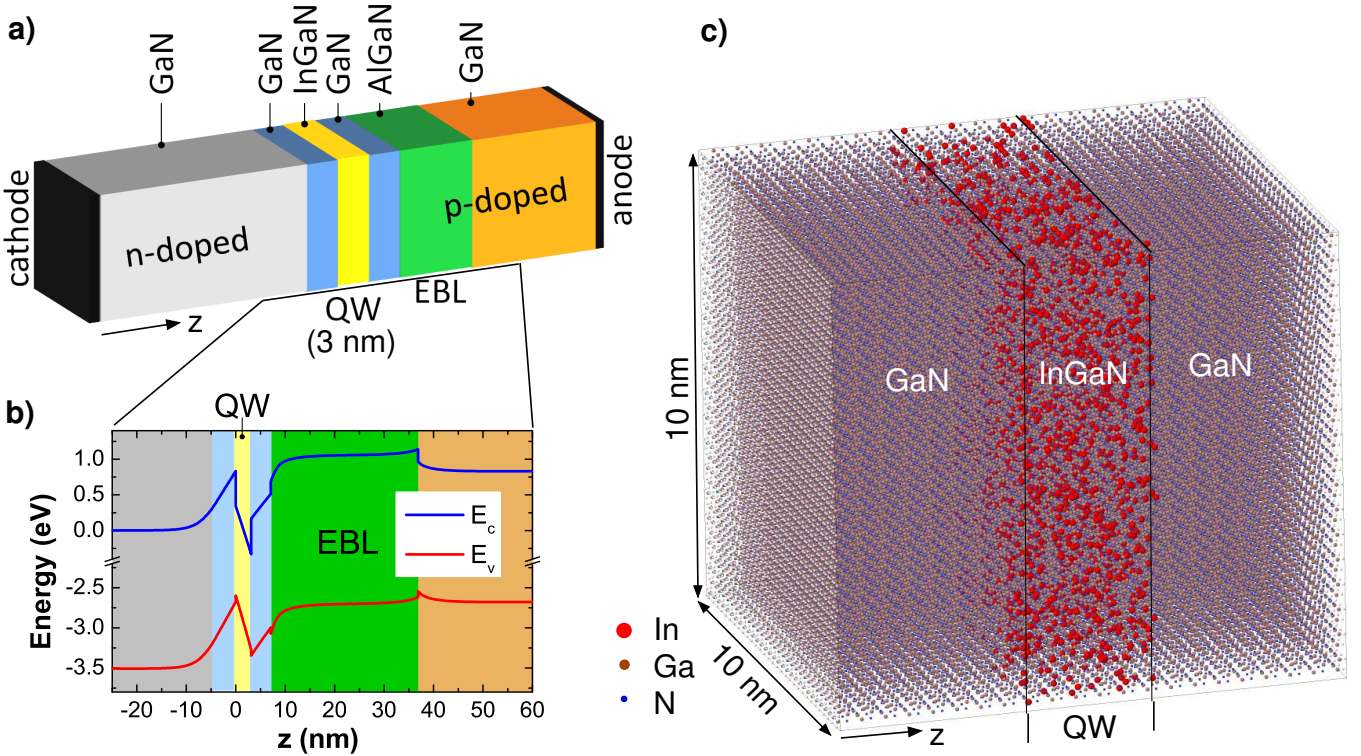}
\caption{{\bf Schematic structure of the simulated devices.} (a) Schematic structure of the SQW LEDs used in the simulations. (b) Typical conduction and valence band edge profile near the predicted maximum IQE point (for 20\% Indium). (c) Atomistic structure of one random alloy sample. The red dots are the Indium atoms.}
\label{device}
\end{figure}

To quantitatively asses the effect of alloy fluctuations on the maximum LED efficiency, we calculated the spontaneous emission properties of c-plane InGaN/GaN single quantum well (SQW) LEDs with Indium contents between 15\% and 35\% and extracted the radiative recombination parameter $B$. This number has then been compared with experimentally determined values based on the ABC model~\cite{Karpov_ABC2014,Schiavon}, which describes the total recombination rate as $R=An+Bn^2+Cn^3$, where $A$ is the defect related Shockley-Read-Hall (SRH) recombination parameter, $n$ is the carrier density, and the parameter $C$ is usually interpreted as the Auger recombination coefficient.
In this model, the internal quantum efficiency (IQE) is given by the ratio of radiative to total recombination, i.e. $\mathrm{IQE}=Bn^2/(An+Bn^2+Cn^3)$.
To obtain $B$, we calculated the confined electron and hole states and the corresponding optical transitions' momentum matrix elements (MME), using an atomistic empirical tight-binding (ETB) approach~\cite{jancu98} and assuming an operating point of the LED near the maximum IQE.
We performed the ETB calculations using both a homogeneous effective medium approximation (virtual crystal approximation) and a random alloy approach in order to quantify the effect of alloy fluctuations in the InGaN.
For the construction of the atomic structure we assumed a uniform random alloy, since recent experimental evidence suggests that InGaN layers of good quality should not deviate much from the ideal uniform case~\cite{Kret_2007,Galtrey_2008,Silvija_2013}.
The random alloy calculations have been performed on 30 random structures for each Indium content.
The schematic device structure used for the simulations, a typical band diagram and a random atomistic structure are shown in Fig.~\ref{device} for the case of 20\% Indium.

\begin{figure}[hbt]
\centering
  \subfloat[]{\includegraphics[width=0.5\columnwidth]{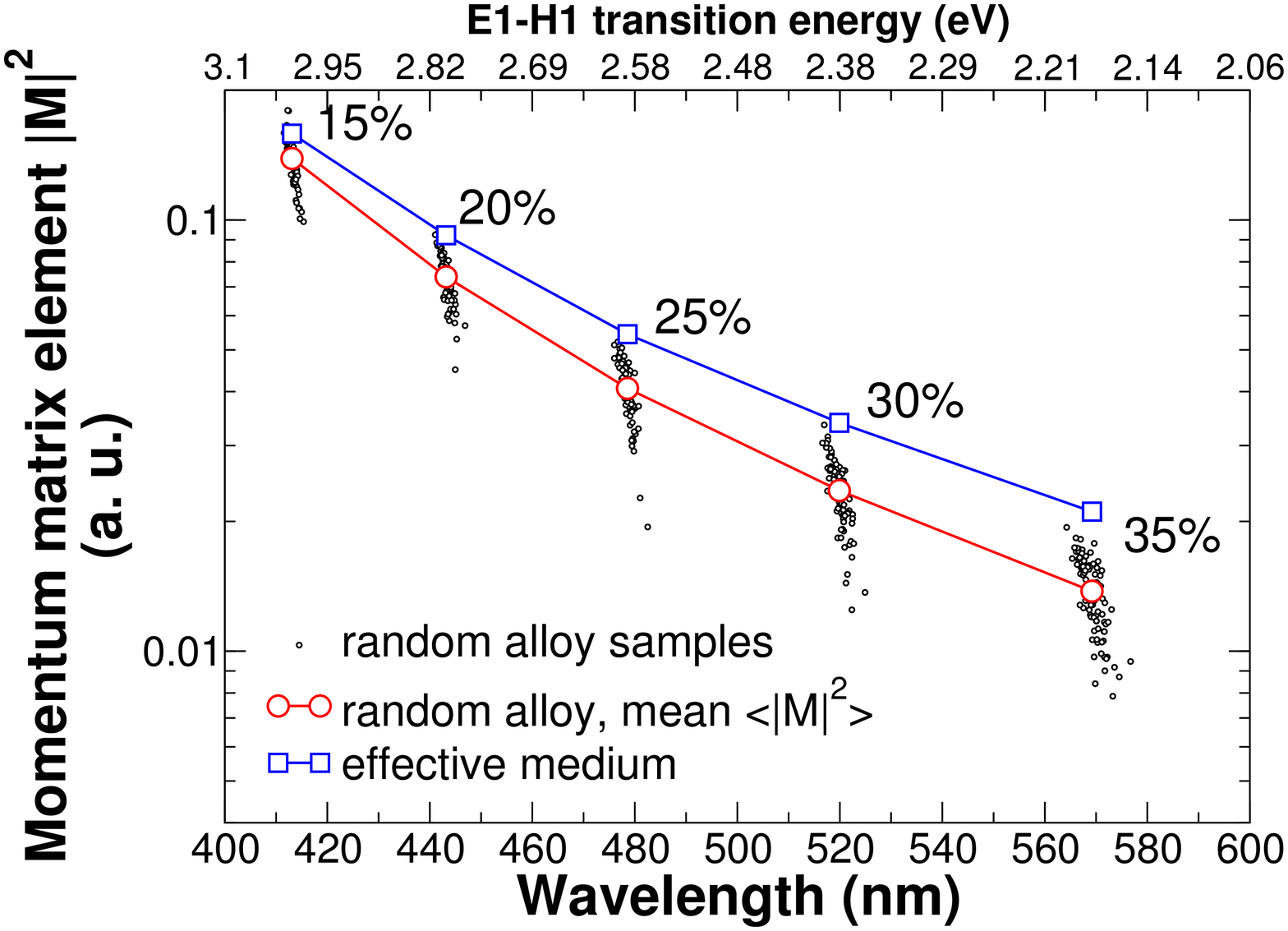}\label{simulation:mme}}
  \hfill
  \subfloat[]{\includegraphics[width=0.5\columnwidth]{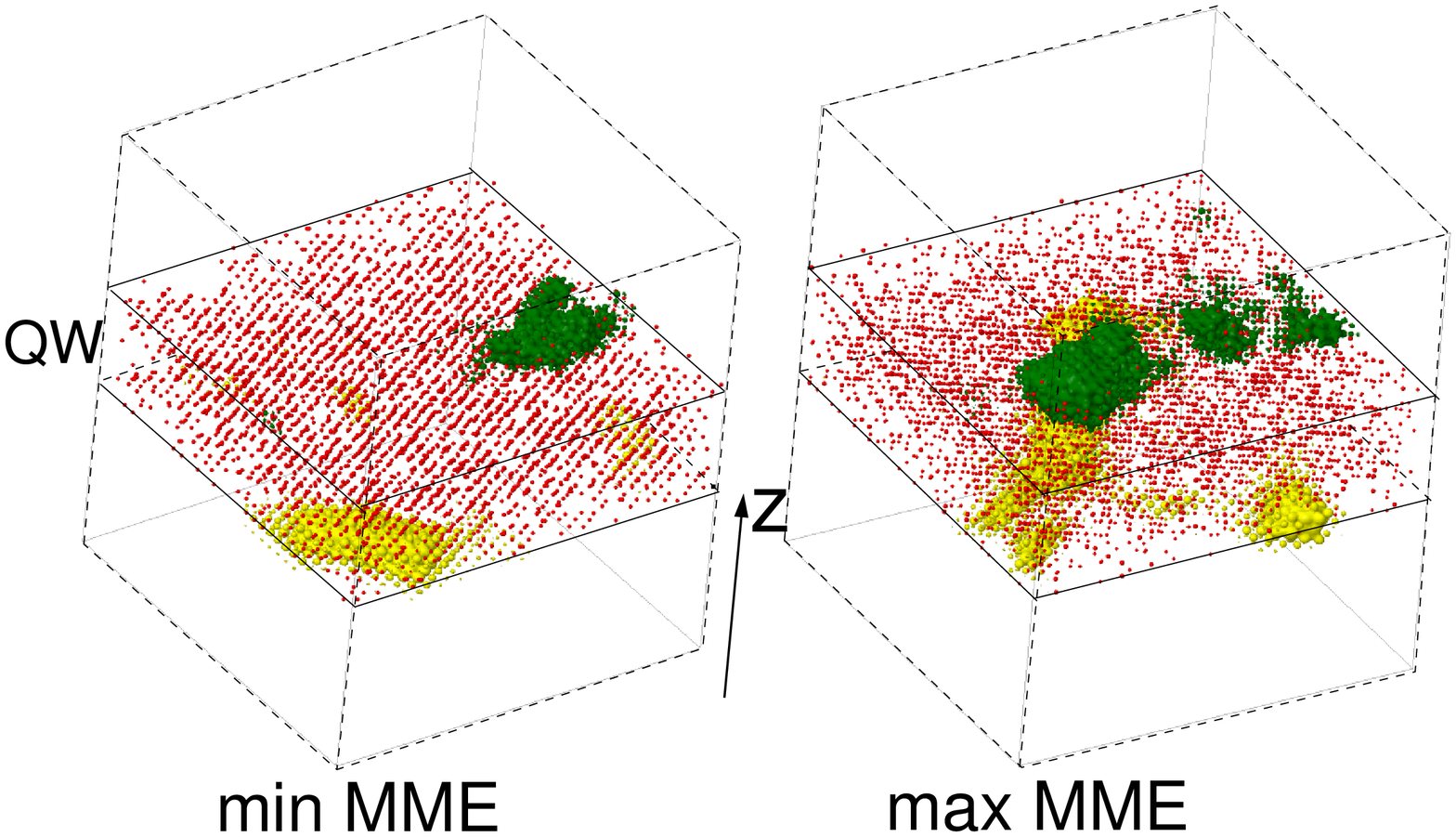}\label{simulation:states}}
  \caption{{\bf Ground state momentum matrix elements and wavefunctions.} (a) Ground state transition matrix elements in atomic units calculated at zero in-plane momentum ($\Gamma$-point) using random alloy and effective medium approach, respectively. 
	  (b) Ground state electron (green) and hole (yellow) wavefunctions for the 30\% In QW with smallest and largest momentum matrix element. The isosurfaces containing 10\% of the total ground state density are shown. The red dots are the Indium atoms.}
\label{simulation}
\end{figure}

Figure~\ref{simulation:mme} shows a scatter plot of the ground state MME for zero transverse momentum (i.e. at the $\Gamma$-point) for the different random samples in comparison with the values obtained in effective medium approximation.
We observe a linear correlation between ground state transition energy and MME, and an increasing spectral spread, confirming earlier results~\cite{Lopez_2014,Caro2013,Schulz2014,AufderMaur2014_rnd}.
The MMEs obtained assuming an effective medium are strictly higher than the random alloy values, indicating that simulations based on this assumption tend to overestimate spontaneous emission strength.
The scattering of the MME can be attributed to variations in wavefunction overlap in the quantum well plane due to Indium fluctuations~\cite{AufderMaur2014_rnd}, as shown in Fig.~\ref{simulation:states}, where the ground state electron and hole wavefunctions are shown for the 30\% Indium QW having smallest and largest MME, respectively.
The strong spread in the values of the MME is due to the fact, that the lateral fluctuations of the electron and hole wavefunctions are largely independent, since they are subject to alloy fluctuations on distant atomic planes due to spatial separation along the crystal's c-axis.

For a direct comparison with experimental data we estimated the radiative recombination coefficient $B$ from the simulation results.
From the statistical ensembles for each Indium content we calculated the mean spontaneous recombination rate $\bar{R}_\mathrm{sp} = 1/N\sum_iR_{\mathrm{sp},i}$ and similarly the mean electron and hole densities $\bar{n}$ and $\bar{p}$, where $R_{\mathrm{sp},i}$ are the spontaneous recombination rates for each random sample calculated by Fermi's golden rule~\cite{chuang_book}, and $N$ is the ensemble size.
Then we extracted an effective radiative recombination parameter $B_{\mathrm{eff}}$ defined such that $\bar{R}_\mathrm{sp} = B_{\mathrm{eff}}\bar{n}\bar{p}$.
This is based on the assumption that the macroscopically observed recombination rates and carrier densities are the spatial mean values.

\begin{figure}[hbt]
\centering
\subfloat[]{\includegraphics[scale=0.35]{IQE.eps}\label{iqe_curves}}
\hfill
\subfloat[]{\includegraphics[scale=0.35]{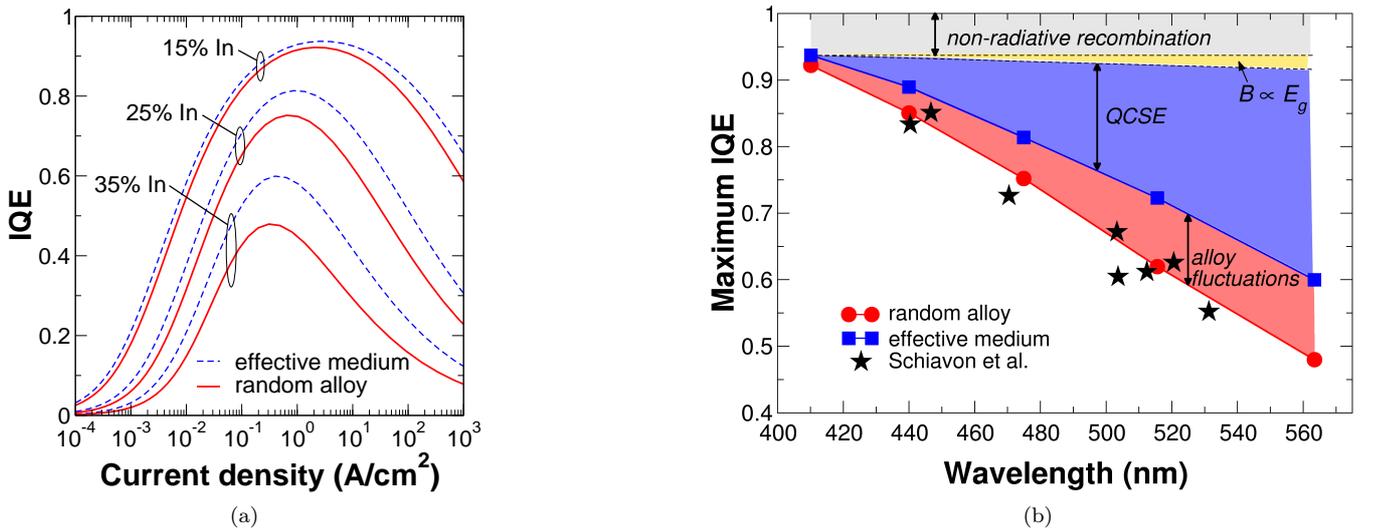}\label{maxIQE}}
\caption{{\bf Comparison of IQE between simulation and measurement.} (a) Simulated IQE versus current density curves, showing the effect of random alloy fluctuations with respect to the effective medium approximation. (b) Wavelength dependency of the peak IQE as expected from simulations compared with the value obtained using the measured $B$. For the SRH and Auger rate coefficients we have assumed constant values of $A=7\cdot10^5$ s$^{-1}$ and $C=10^{-31}$ cm$^{6}$s$^{-1}$, respectively~\cite{Schiavon}.
Random alloy fluctuations induce an additional drop in IQE of up to 0.1.}
\label{IQE}
\end{figure}

Based on the ABC model and using the experimentally extracted recombination coefficients we can estimate the effect of random alloy fluctuations on the maximum IQE that can be expected from a c-plane InGaN/GaN SQW LED, and in particular its wavelength dependence.
In the ABC model the maximum IQE is given by $\mathrm{IQE}_{\mathrm{max}}=B/(B+2\sqrt{AC})$.
For the estimate we use constant $A$ and $C$ parameters, taking the SRH recombination coefficient $A=7\cdot10^5$ s$^{-1}$ from~\cite{Schiavon} and the approximately constant Auger coefficient $C=10^{-31}$ cm$^{6}$s$^{-1}$ from the same source.
This allows to compare the wavelength dependence of the maximum IQE under the best-case assumption of constant non-radiative recombination.
Fig.\ref{iqe_curves} shows the calculated IQE as a function of current density.
It can be seen that uniform random alloy fluctuations lead to an additional progressive reduction of the IQE of up to roughly 0.1 with respect to the prediction based on an effective medium approximation.
The trend with wavelength of the peak IQE predicted by the random alloy calculation is in good agreement with the experimental data, as shown in Fig.~\ref{maxIQE}.
We note, that assuming some non-uniformity in the Indium distribution does not significantly change the calculated $B$, whereas it substantially broadens the predicted emission spectra (see Supplementary material).
Therefore, uniform random alloy fluctuations can be identified to give an important contribution to the ``green gap'', and our results allow to quantify this contribution to account for up to 30\% in the green spectral region.
Fig.~\ref{IQE} also shows, that the combination of non-radiative recombination, dependency of optical strength on emission energy, QCSE due to the internal field and finally alloy fluctuations may comprehensively explain the origins of the green gap in nitride LEDs.

In summary, based on atomistic simulations of c-plane InGaN/GaN SQW LEDs including uniform random alloy fluctuations we predicted the radiative recombination parameter $B$ for different mean Indium contents.
We have shown that the wavelength dependence of $B$ is compatible with experimental findings, and that simulation approaches based on homogeneous effective media approximations, typically used for device simulations, overestimate $B$ by an amount proportional to the mean Indium content in the QW.
Comparing the predicted maximum IQE with the one obtained using measured values of $B$, assuming for the non-radiative recombination parameters wavelength independent values, leads to the conclusion that alloy fluctuations give rise to an important material intrinsic contribution to the ``green gap'', which in the studied structures can be as big as 30\% of the total IQE drop at green wavelengths.
Since the strong scattering of the momentum matrix elements is due to the internal field induced uncorrelated lateral fluctuations of electron and hole wavefunctions, it can be expected that in non-polar QWs the effect of the random alloy fluctuations would be considerably reduced due to the absence of QCSE.

\section*{Methods}

For each Indium composition, we first solved self-consistently the Schr\"odinger/drift-diffusion equations for a bias voltage approximately resulting in the maximum IQE, using 8-band k$\cdot$p based envelope function approximation for the electronic states~\cite{chuang2}.
Then we mapped the obtained electrostatic potential onto random atomistic structures, which have been built by randomly substituting Gallium atoms with Indium atoms in the QW such as to reproduce a uniform random alloy, keeping the total number of substituted atoms fixed in order to control the mean Indium composition.
For the atomistic structures we assumed a periodic supercell in the QW plane of 10$\times$10 nm$^2$, resulting in structures containing roughly 100,000 atoms.
The atomic positions in the atomistic structure have been relaxed using a modified Keating's valence force field (VFF) model~\cite{niquet_vff}.
Then we calculated the first four spin degenerate electron and hole states from an empirical tight-binding (ETB) hamiltonian, employing an sp$^3$d$^5$s$^*$ parametrization~\cite{jancu} and using a GPU accelerated Jacobi-Davidson algorithm~\cite{Rodrigues_JCEL_2015}.
From these states we obtained the momentum matrix elements, spontaneous emission spectra and the carrier densities, using Fermi-Dirac statistics and fixed quasi Fermi levels.
Integrations in k-space have been approximated by summing over 4 k-points of the reduced Brillouin zone, since the large periodicity of 10 nm is leading to a very small Brillouin zone.
All simulations have been performed using the tiberCAD multiscale simulation software~\cite{AufderMaurTED2011}.

\section*{Acknowledgment}
The research leading to these results has received funding from the European FP7 Project NEWLED under grant agreement N. 318388.

\section*{Authorship}
M.A., A.P. and A.D. conceived the study. M.A. set up and performed the simulations, analysed the data and wrote the manuscript. M.A., A.P. and G.P. contributed to the software implementation of the models, W.R. parallelized the tight-binding code. All authors discussed the results and edited the manuscript.



\bibliographystyle{unsrt}
\bibliography{biblio}

\end{document}